\documentclass
[aps,prd,onecolumn,superscriptaddress,amsfont,graphicx,nofootinbib]{revtex4}%
\usepackage{color,graphicx,epsfig}
\usepackage{ifpdf}
\usepackage{amsmath}
\usepackage{bm}
\usepackage{color}
\usepackage[english]{babel}
\usepackage{graphicx}%
\usepackage{amsfonts}%
\usepackage{amssymb}
\usepackage{braket}

\newcommand{\WA}{\scriptscriptstyle\rm WA}

\definecolor{Red}{rgb}{1.,0.,0.}

\begin{document}

\title{Lepton energy moments in semileptonic charm decays}

\author{Paolo Gambino} 
\affiliation{Dip. Fisica Teorica, Univ. di Torino, \& INFN Torino, I-10125 Torino, Italy}
\email[Electronic address:]{gambino@to.infn.it} 

\author{Jernej F. Kamenik}
\email[Electronic address:]{jernej.kamenik@ijs.si} 
\affiliation{J. Stefan Institute, Jamova 39, P. O. Box 3000, 1001  Ljubljana, Slovenia}

\date{\today}

\begin{abstract}
We search for  signals of Weak Annihilation in inclusive  semileptonic $D$ decays. We consider both the widths and the lepton energy moments, which are quite sensitive probes. Our analysis of Cleo data shows no clear evidence of Weak Annihilation, and allows us to put bounds on their relevance in  charmless $B$ semileptonic decays.
\end{abstract}

\maketitle

\section{Introduction}

The value of $|V_{ub}|$ preferred by current global analyses of CKM  data is about 15\% smaller than the one extracted from inclusive charmless semileptonic $B$ decays~\cite{UTfit}. Though not very significant, the discrepancy has prompted a reexamination of the sources of theoretical uncertainty in the inclusive determination \cite{ckmbook}. Weak Annihilation (WA) contributions are generally considered an important source of uncertainty in the Operator Product Expansion (OPE)  that describes the inclusive $B$ decays \cite{OPE}, and affect especially  the high $q^2$ and lepton endpoint analyses.
They appear in the OPE as $1/m_b^3$ corrections involving the matrix elements of dimension-6 four-quark operators, and affect both  the total $B \to X_u \ell \bar\nu$ 
decay rate and the charged lepton energy spectrum
\cite{Bigi:1993bh,Bigi:1997dn}.

Early estimates of the relevant matrix elements were derived in the framework of QCD sum rules~\cite{QCDSRHQ}. They were also computed on the lattice in the static heavy quark limit \cite{DiPierro:1998ty} and with propagating heavy quarks \cite{Becirevic:2001fy}. However, it was soon realized  \cite{Pirjol:1998ur, Uraltsev:1999rr}
 that the light flavour component of the WA operators can differ from the light valence quark of the $B$-meson leading to the so-called non-valence WA contributions which are extremely difficult to study non-perturbatively. In addition, contrary to valence WA contributions, the latter cannot be constrained by comparing charged and neutral $B$ meson semileptonic decay rates and a common prejudice that the former should dominate may be unfounded.

It was also noted in ref.\cite{Bigi:1993ey,Uraltsev:1999rr,Voloshin:2001xi} that the WA matrix elements that enter $B \to X_u \ell \bar\nu$ decay can be constrained via the semileptonic decays of $D$ and $D_s$ mesons, using heavy quark symmetry. Several authors have attempted to extract information on WA contributions from the measured total semileptonic rates of $D^{0,+}$ and $D_s$, most recently in \cite{Becirevic:2008us, Bigi:2009ym, Ligeti:2010vd}. For instance, one may attribute the observed differences in $D^{+,0}$ and $D_s$ semileptonic widths \cite{:2009pu}
\begin{equation}\label{eq:cleo-c}
\Gamma(D^+\to X e^+\nu)/\Gamma(D^0\to X e^+ \nu)=0.985(28) \ ,  \ \ \
\Gamma(D^+_s\to X e^+\nu)/\Gamma(D^0\to X e^+ \nu)=0.828(57)
\end{equation}
 to the valence spectator quark WA contributions in $D_s$ decays, since they are Cabibbo suppressed in the $D^+$ case and  completely absent in $D^0$ decays. However, additional contributions to this difference arise from $SU(3)$ breaking in the matrix elements of 
all dimension 5 and 6 operators, that contribute  significantly to the total rates \cite{Bigi:1993ey}.  
Moreover, the slow convergence of the $1/m_c$ expansion, the generally large perturbative corrections to the semileptonic charm width, and the strong dependence on the charm mass may obscure the determination of the non-valence WA component from the widths \footnote{For a related discussion on the charm meson lifetimes see \cite{Bigi:1993ey}.}.

Our strategy in this paper is  to consider also the moments of the lepton energy spectra 
in the OPE and compare them with recent experimental results from Cleo~\cite{:2009pu}. Not only are the moments free of the 
strong  dependence on the charm quark mass and its associated uncertainty, but their perturbative and non-perturbative corrections tend to cancel as well. Moreover, since  WA is expected to dominate the spectrum endpoint,  leptonic moments might be more sensitive to WA contributions than the total rate.  A study of moments in $D$ semileptonic decays is also instrumental for a critical reassessment of the OPE in charm decays many years after 
\cite{Blok:1994cd}, 
in view of the recent experimental results and of the successful application to $B$ semileptonic decays.

In the next Section, we present the experimental results on the leptonic spectra and compute from them the first leptonic moments. The  OPE calculation of the moments is presented in Section III, while we derive and discuss our results in Section IV. We conclude with a brief summary.

\section{Experimental data}

Recently the Cleo Collaboration has measured the electron spectra of inclusive semileptonic $D^{+,0}$ and also $D_s$ decays with a lower cut on the electron momentum in the lab frame of $p_e > 0.2$ GeV \cite{:2009pu}. They extract the total decay rates by extrapolating the spectra over the remaining phase-space, using a theoretically modeled sum over known exclusive modes. Unfortunately, they do not provide higher leptonic energy moments. 

\begin{figure}
  \includegraphics[height=.24\textheight]{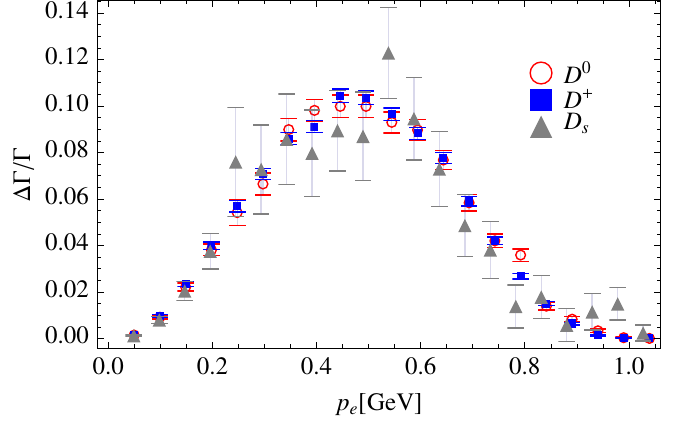}
  \caption{\label{fig:spectra} Normalized charm semileptonic decay spectra \cite{:2009pu}, extrapolated to $p_e=0$ and boosted to the reference frame of the decaying particle.}
\end{figure}

Cleo give the electron energy spectra in the laboratory frame. Since the decaying $D^{0,+}$ mesons are pair-produced with center-of-mass energy $E_{CM} = 3.774$ GeV,  they are boosted in this frame with $\beta = 0.14, 0.15$, respectively.  In the case of $D_s$ mesons,
the situation is more complicated since roughly half of them are produced in the primary vertex at $E_{CM}=4.170$ GeV associated with $D_s^*$ mesons ($\beta = 0.2$). The other half then comes from radiative and hadronic $D^*_s$ decays. 

We extract the moments without cuts by first extrapolating the measured spectra towards $p_e=0$. We use a procedure similar  to the one employed by Cleo themselves, but instead of using a sum over exclusive modes with model form factors, we use the  theoretical  form of the spectrum in the $p_e\approx 0$ region, where the OPE is expected to provide a satisfactory  description. In particular, for $p_e\to 0$ the spectrum must die at least like $p_e^2$. Thus we fit the first four measured bins to  $d \Gamma / d x = a x^2 (1+ b x)(1-x)$, with $x=2E_e/m_c$. In the case of $D_s$, due to the larger statistical uncertainties, we set $b=0$ and fit only to the first two measured bins. In the fits we take the reported systematic errors to be fully correlated between different bins. 
 Using the fitted formulas we have computed the rates and obtained the following branching fractions 
\begin{align}
\mathcal B (D^0\to X e \nu) &= 0.064(1)  \,, & \mathcal B (D^+\to X e \nu) &=  0.161(3)\,, & \mathcal B (D_s\to X e \nu) &= 0.065(4) \,,
\label{eq:br}
\end{align}
fully compatible with Cleo reported results. We then compute in the lab frame the first two leptonic energy moments normalized to the total rate, again assuming the systematic errors to be fully correlated among different bins  and obtain
\begin{align}
\braket{E_e}_{lab}^{D^0}  &= 0.465(3) \mathrm{GeV} \,, & \braket{E_e^2}_{lab}^{D^0} &= 0.248(2) \mathrm{GeV}^2\,, \\   
\braket{E_e}_{lab}^{D^+}  &= 0.459(1) \mathrm{GeV} \,, & \braket{E_e^2}_{lab}^{D^+}&= 0.242(1) \mathrm{GeV}^2\,, \\   
\braket{E_e}_{lab}^{D_s}  &= 0.466(12) \mathrm{GeV} \,,& \braket{E^2_e}_{lab}^{D_s} &= 0.254(13) \mathrm{GeV}^2  \,.
\label{eq:lab}
\end{align}
The dependence of these values upon our extrapolation ansatz is expected to be smaller than the corresponding effect in the total rate, since the contribution  of the $p_e\sim 0$ region to the $n$-th (unnormalized) moment is $(2 p_e/m_D)^n$ suppressed compared to its contribution to the total rate.  Thanks to the correlations between numerators and denominators,  the total errors in (\ref{eq:lab}) are significantly smaller than in (\ref{eq:br}).

In order to compare these values to theoretical predictions in the $D$ meson frame one still needs to take into account the boost factors. The energy of the final state electron in the lab frame is given by $E'_e = \gamma E_e (1-\beta \cos\theta)$, where 
$E_e$ is the electron energy in the $D$ meson frame and $\theta$ is the angle between 
the momentum of the electron and  the one of the $D$ meson in the lab frame. 
We thus obtain that $\braket{E'_e} \equiv \braket{E_e}_{lab} = \gamma \braket{E_e}$ with $\gamma = 1.009, 1.012$ for Cleo's $D^{+,0}$ mesons, respectively. Similarly, for the $49\%$ of $D_s$'s which come from the primary vertex, $\gamma\simeq1.02$. 
The remaining $51\%$ receive an additional boost in an arbitrary direction since they originate from $D_s^*$ decays $D_s^*\to D_s \gamma$ ($94\%$) or $D^*_s\to D_s \pi$ ($6\%$).
In view of the precision in (\ref{eq:lab}), the effect of this additional boost is always negligible. 
The directional averaging for the second electron energy moment $\braket{E_e^2}$ results in $\braket{{E'}^{2}_e} = \braket{ \gamma^2 (1+\beta^2/3)}\braket{E^2_e}$ which again can be readily computed for all cases. The extrapolated rest-frame spectra are displayed in 
Fig.~\ref{fig:spectra}. Our results for the moments in the $D$ mesons rest frame are
\begin{align}
  \braket{E_\ell}^{D^0}_{exp} & = 0.459(3) \mathrm{GeV}\,, &    \braket{E_\ell^2}^{D^0}_{exp} & = 0.240(2) \mathrm{GeV}^2\,, &  \braket{(E_\ell- \braket{E_\ell})^2}^{D^0}_{exp} & = 0.029(2) \mathrm{GeV}^2\,,\nonumber\\
  \braket{E_\ell}^{D^+}_{exp} & = 0.455(1) \mathrm{GeV}\,, &    \braket{E_\ell^2}^{D^+}_{exp} & = 0.236(1) \mathrm{GeV}^2\,, &  \braket{(E_\ell- \braket{E_\ell})^2}^{D^+}_{exp} & = 0.029(1) \mathrm{GeV}^2\,,\\
  \braket{E_\ell}^{D_s}_{exp} & = 0.456(11) \mathrm{GeV}\,, &    \braket{E_\ell^2}^{D_s}_{exp} & = 0.239(12) \mathrm{GeV}^2\,, &  \braket{(E_\ell - \braket{E_\ell})^2}^{D_s}_{exp} & = 0.031(12) \mathrm{GeV}^2\,. \nonumber
\end{align}
Here we have also listed the variances ($\sigma_E^2$) of the spectra.  Within the stated uncertainties there is no sign of a difference between the moments of $D_s$ and $D^{0,\pm}$. This is at odds with what one would naively  expect  from  eq.(\ref{eq:cleo-c}).
However, as mentioned in the Introduction,  the concurring contribution of SU(3) violation in the matrix elements of higher dimensional operators might provide a partial explanation.

 One can estimate the order of magnitude of  $SU(3)$ breaking corrections by comparing the hyperfine splitting of $D^{0,+}$ and $D_s$ mesons, 
$\Delta_{D_q}^{hf}=3 (m_{D_q^*}^2-m_{D_q}^2)/4$ (which is related to the OPE parameter $\mu_G^2$):
\begin{equation}\label{eq:muG}
\Delta^{hf}_{D^+} = 0.409(1) \mathrm{GeV}^2\,, \qquad \Delta^{hf}_{D^0} = 0.413(1) \mathrm{GeV}^2\,, \qquad \Delta^{hf}_{D_s} = 0.440(2) \mathrm{GeV}^2\,.
\end{equation}
 We see that even isospin violation between $D^{+,0}$ mesons is manifest, at the expected $1\%$ level.
  On the other hand $SU(3)$ violation is sizable and of the order $10\%$. Lattice QCD studies of $f_D$ and $f_{D_s}$ \cite{fD} suggest that even  20\% violation can be expected in certain quantities.
 Since  the $1/m_c^2$ and $1/m_c^3$ contributions  to the total semileptonic rate can be as large as $50\%$ of the leading order estimate (depending on the charm quark scheme and scale),  and since one can expect  $SU(3)$ violation of similar size in the matrix elements of the relevant power suppressed operators, an $O(10\%)$ $SU(3)$ breaking in the widths is not unlikely.
 In order to reliably extract possible WA contributions from the measured total semileptonic rates of $D$ and $D_s$ mesons \cite{:2009pu}, one would therefore need to  estimate the size of $SU(3)$ violation in all these matrix elements. In the case of normalized moments, on the other hand,  some of the leading power corrections cancel out  and one might be more directly sensitive to WA contributions.

\section{Leptonic moments in the OPE}
The perturbative corrections to moments of some kinematic distributions 
in semileptonic $b$ decays are now known through $O(\alpha_s^2)$ \cite{Pak:2008qt}.
In the case of semileptonic charm decays, however,  only the $O(\alpha_s)$ \cite{Czarnecki:1994pu}
and  $O(\beta_0\alpha_s^2)$ corrections are readily 
available using  \cite{Aquila:2005hq}. For what concerns the power corrections, 
explicit expressions for the leptonic spectrum  at $O(1/m_c^2)$ can be found in the last paper of \cite{OPE}, while we have computed the $O(1/m_c^3)$ contributions from the form factors given in \cite{Gremm:1996df}. We neglect $O(1/m_c^4)$ corrections
\cite{Dassinger:2006md} and $O(\alpha_s/m_c^2)$ corrections, for which only the $\mu^2_\pi$ contribution to the rate is known.

We calculate the semileptonic rate and the total moments from the lepton energy spectra in terms of $x \equiv 2 E_e / m_c$, where $E_e$ is the electron energy in the $D$ meson inertial frame. For simplicity, we only consider total leptonic moments, without lower cuts on the lepton energy, and neglect the lepton mass. We define
\begin{eqnarray}
\Gamma^{(n)} \equiv \int_0^{(1- r)}  \frac{d\Gamma}{dx} x^n d x &=&  \frac{G_F^2 m_c^5}{192 \pi^3} |V_{cs}|^2 \left[ f^{(n)}_0(r) + \frac{\alpha_s}{\pi} f^{(n)}_1(r) +  \frac{\alpha_s^2}{\pi^2} f^{(n)}_2(r) +  \frac{\mu_\pi^2}{m_c^2} f^{(n)}_\pi (r) +  \frac{\mu_G^2}{m_c^2} f^{(n)}_G (r) \right. \nonumber\\
&&\left.+\frac{\rho_{LS}^3}{m_c^3} f^{(n)}_{LS} (r)+ \frac{\rho_D^3}{m_c^3} f^{(n)}_D (r) +  \frac{32 \pi^2}{ m_c^3} %(1-r)^n 
B_{\WA}^{(n)s} \right]\,,
\end{eqnarray}
where $r=m_s^2/m_c^2$, $\alpha_s\equiv \alpha_s(m_c)$, and $\mu^2_{\pi,G}$, 
$\rho^3_{D,LS}$ are the $D$ meson matrix elements of the dimension 5 and 6 local operators appearing in the OPE.
$B_{\WA}^{(n)s}$ is the WA contribution to the $n$-moment: as we will explain in a moment, it is related to the matrix elements of four-quark operators.
In addition, 
there are also Cabibbo suppressed contributions, which can be  included in the analysis by using the above formula in the limit $r\to 0$ and by replacing $V_{cs}$ with $V_{cd}$ and similarly $B_{\WA}^{(n),s}$ with $B_{\WA}^{(n),d}$. They contribute to the total rate at the  level of 5\%, but their effect is highly suppressed in the normalized moments, with the possible exception of WA contributions. 

The tree-level spectrum in charm semileptonic decays is softer than in the analogous $b$ decays and peaks at $x\approx 0.6$, a feature qualitatively evident already in Fig.~1.
The lowest order expressions for the rate and the first two moments read 
\begin{align}
f_0^{(0)}&= 1-8 r + 8 r^3 - r^4 - 12 r^2 \log r\,, \nonumber \\
f_0^{(1)} &= [2 r^5 - 15 r^4 + 60 r^3 - 20 r^2 -60 r^2 \log r - 30 r + 3]/5\,, \nonumber\\
f_0^{(2)} &= [-r^6 +8r^5-30r^4+80r^3-35r^2-60r^2\log r - 24 r + 2]/5\,. \nonumber
\end{align}
The associated $O(\alpha_s)$ corrections are known in closed form only for the total rate. We employ accurate numerical  interpolations valid in the range $0.04<\sqrt r < 0.2$ in the pole mass scheme
\begin{align}
f_1^{(0)} (r)& = 2.86 \sqrt r - 3.84 r \log r -2.47\,, \nonumber\\
f_1^{(1)}(r) &= 2.11 \sqrt r - 2.43 r  \log r - 1.56\,, \nonumber\\
f_1^{(2)}(r) &= 1.66 \sqrt r  - 1.71  r \log r - 1.09\,. \nonumber
\end{align}
Similarly, while there is an analytic result for $f^{(0)}_2(r)$ as an expansion in $r$ \cite{Pak:2008qt}, we employ only the  BLM approximation \cite{Aquila:2005hq} in the form of numerical interpolations in the pole mass scheme:
\begin{align}
f^{(0)}_2(r) &= \beta_0 [8.16 \sqrt r - 1.21 r \log r - 3.38]  \,, \nonumber\\
f^{(1)}_{2}(r) &= \beta_0  [6.15 \sqrt r - 0.35 r \log r - 2.24]\,, \nonumber\\
f^{(2)}_{2}(r) &= \beta_0  [4.99 \sqrt r + 0.11 r \log r - 1.64]\,, \nonumber 
\end{align}
where $\beta_0$ is the QCD beta function $\beta_0 = 11- 2 n_f /3$ and $n_f=3$ is the number of light flavors. 
The above expressions can be translated to any other scheme (and associated IR scale) using the condition that the decay rate be scheme (and scale) independent at each perturbative order together with the known scale dependence of the heavy quark mass and power suppressed non-perturbative parameters.
For the leading power corrections we get
 \begin{align}
 f^{(0)}_\pi (r)&= -f^{(0)}_0(r)/2 \,, \quad\quad\quad\quad  f^{(1)}_\pi (r)= 0 \,, \quad\quad\quad\quad
 f^{(2)}_\pi (r) = \frac56 f^{(2)}_0(r) \,, \nonumber\\ 
 f^{(0)}_G(r) &= \frac12 f^{(0)}_0(r) - 2 (1-r)^4\,, \quad
 f^{(1)}_G(r) =  \frac13 f^{(1)}_0(r) - \frac65 (1-r)^5\,,\quad
 f^{(2)}_G(r) = \frac16 f^{(2)}_0(r) - \frac45 (1-r)^6\,,\nonumber\\ 
 f^{(0)}_{LS}(r) &= -f^{(0)}_G(r)\,,\quad\quad\quad %-f^{(0)}_0/2 + 2 (1-r)^4\ \ \ \ \
 f^{(1)}_{LS}(r) = \frac85 (1-r)^5\,,\quad\quad\quad
 f^{(2)}_{LS}(r) = \frac56 f^{(2)}_0(r)+\frac43 (1-r)^6\,,\nonumber\\ 
 f^{(0)}_{D}(r) &= \frac{77}6 + O(r) %-\frac{44}3 r 
 + 8\log \frac{\mu_{\WA}^2}{m_c^2} \nonumber
 \,, \quad
 f^{(1)}_{D}(r) =\frac{78}{5}+ O(r) + 8\log \frac{\mu_{\WA}^2}{m_c^2} 
 \,,\nonumber\\ 
 f^{(2)}_{D}(r) &= \frac{87}5 + O(r)+ 8\log \frac{\mu_{\WA}^2}{m_c^2} 
 \,,\nonumber
\end{align}
where $\mu_{\WA}$ is the $\overline{\rm MS}$ renormalization scale associated to 
the mixing of the Darwin and WA operators, an $O(\alpha_s^0)$ effect \cite{bumom,Vub}.
The $\mu_{\WA}$ dependence 
cancels against the implicit  scale dependence of $B_{\WA}^{(n),q}(\mu_{\WA})$. A change of $\mu_{\WA}$ therefore shifts part of the Darwin operator contribution into the WA contribution, with important consequences for the error analysis.
We find that for  values of $\mu_{\WA}$  below 1GeV the size of the $\rho^3_D$ coefficient in the width is comparable to that of  other power corrections. 

 WA contributions can be identified with the matrix elements of the dimension-6 four quark operators 
 $$\mathcal O^{Qq}_{1} = \bar Q \gamma_\mu (1-\gamma_5) q \,\bar q \gamma^\mu (1-\gamma_5) Q \quad \mathrm{ and}\quad \mathcal O^{Qq}_{2} = \bar Q (1-\gamma_5) q \,\bar q  (1-\gamma_5) Q\,,$$
where $Q$ is the heavy quark. The matrix elements that enter the charm meson decay rates  are defined as
\begin{equation}\label{eq:me}
B^{q}_{\WA} = \frac{1}{2m_D} \bra{D} \mathcal O^{q}_2 - \mathcal O^{q}_1 \ket{D} = \frac{1}{2} m_D f_D^2 (B^{q}_{D,2}-B^{q}_{D,1})\,,
\end{equation}
where $B^{q}_{D,i}$ parameterize the deviation from the factorization approximation: $B^{q}_{\WA}$ vanishes in the limit of factorization.
As it was recognized long ago \cite{Bigi:1993bh}, WA  is 
localized at  the endpoint of the lepton energy spectrum, and can be approximately expressed by a delta function at the partonic endpoint. In this case, we would expect
$B_{\WA}^{(n),q}= B_{\WA}^{q}$, for all $n$, up to small $O(r)$ effects.
In fact, gluon bremsstrahlung and hadronization effects  are expected to smear the WA
 contribution over a region of electron energy around the partonic threshold $(m^2_c-m_s^2)/2m_c$ corresponding to $x=1-r$. The size and shape of the smearing may affect the various
integrals  $B_{\WA}^{(n),q}$ differently and cannot be predicted, although we expect a small perturbative tail to emerge away from the endpoint. 
Clearly,   smearing towards smaller $E_e$ tends to suppress the WA contributions to higher lepton energy moments,  in which case we expect $B_{\WA}^{(n+1),q}\lesssim B_{\WA}^{(n),q}$. We also know that the WA contributions involve a $SU(3)$ flavor singlet in the final state, and 
may be confined below the two-pion or $\eta$ thresholds  which, for $D$ decays  and
$m_c=1.4$GeV, correspond to $x\le 1.31$ and $x\le1.22$, respectively. For $D_s$ decays 
the allowed region is slightly larger.
Indeed, the WA distributions spread over a region of electron energies of $O(\Lambda_{QCD})$.
In Fig.~2 we show four possible forms of the WA distribution as a function of $x$, assuming only positivity and the presence of a tail towards lower $x$. Two of the forms displayed stop at the $\eta$ threshold, the other two extend to the two-pion threshold. 

The endpoint behavior of WA can also be parameterized in terms of matrix elements 
of local four-quark operators of higher and higher dimension \cite{Leibovich:2002ys},
which determine the WA distribution. At leading order in the heavy quark expansion, however,
the parameters $B_{\WA}^{(n)q}$, i.e.\ the total moments of the WA distribution,
are all determined by the dimension six term, and therefore
they are all equal to $B_{\WA}^{(0)q}$. The smearing is therefore a power
and $\alpha_s$-suppressed effect in $B_{\WA}^{(n)q}$, but it may be phenomenologically important despite the formal suppression.

We will be  primarily interested in a determination of the leading matrix element, $B_{\WA}^{(0)q}$, i.e.\ the zeroth moment of the WA distribution, from $B_{\WA}^{(n)q}$:  the non-negligible smearing of 
WA may lead to  a model-dependent dilution.
In order to quantify this effect on the parameters $B_{\WA}^{(1)q}$, we have considered a number of distributions like those in  Fig.~2.  In general the dilution is
a moderate effect:  $B_{\WA}^{(1)q}\gtrsim 0.8 \,B_{\WA}^{(0)q}$, $B_{\WA}^{(2)q}\gtrsim0.7\,B_{\WA}^{(0)q}$, and of course for WA distributions localized to the right of the partonic endpoint,  
$B_{\WA}^{(n>0)q}\ge B_{\WA}^{(0)q}$. However, distributions characterized by a longer tail will lead to 
a stronger dilution, which may also be enhanced in the normalized moments.
In the next Section, when we try to extract information on the WA from the leptonic moments,  we will carefully take this effect into account.

\begin{figure}
  \includegraphics[height=.24\textheight]{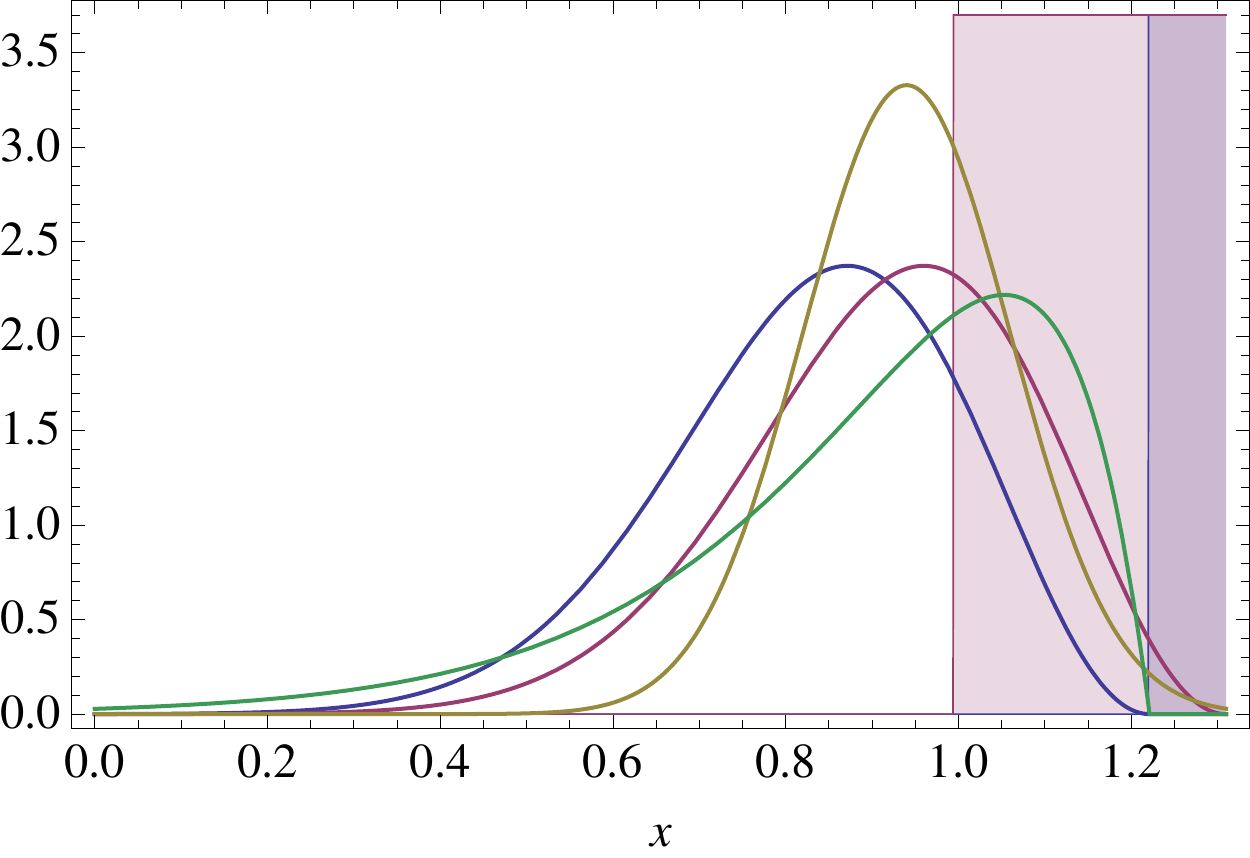}
  \caption{\label{fig:dilution} 
Different distributions of WA as a function of $x=2E_e/m_c$, for $m_c=1.4$GeV.
The distributions extend to $x=1.22$ corresponding to the $\eta$ threshold, or to $x=1.31$, corresponding to the $2\pi$ threshold for $D^{+,0}$.
    }
\end{figure}

The WA contributions to the decays of charmed mesons involving different spectator quarks can be parametrically decomposed as follows:
\begin{eqnarray}
\Gamma^{(n)}_{\WA}(D^0) &\propto& \cos^2\theta_c B^{(n),s}_{\WA}(D^0)  + \sin^2\theta_c B^{(n),d}_{\WA}(D^0) \,, \nonumber\\
\Gamma^{(n)}_{\WA} (D^+) &\propto& \cos^2\theta_c B^{(n),s}_{\WA}(D^+)  + \sin^2\theta_c B^{(n),d}_{\WA}(D^+) \,,\\
\Gamma^{(n)}_{\WA} (D_s) &\propto& \cos^2\theta_c B^{(n),s}_{\WA}(D_s)  + \sin^2\theta_c B^{(n),d}_{\WA}(D_s) \,,\nonumber
\end{eqnarray} 
where $\theta_c$ is the Cabibbo angle.
In the isospin limit, the number of independent contributions is reduced  by the identification $B^{(n),s}_{\WA}(D^0)=B^{(n),s}_{\WA}(D^+)$. Any difference in the moments between the $D^0$ and the $D^+$ can only be due to isospin violation or Cabibbo suppressed contributions. Since the data do not show any significant difference in the rates or the moments of $D^0$ and $D^+$ 
 in what follows we will neglect such Cabibbo suppressed effects. We are then left with just two distinct contributions, which can be identified with the valence and non-valence WA terms involving the $s$ quark. The first one, $B^{(n),s}_{\WA}(D_s)$, only contributes to the $D_s$ decays, while the second one,  $B^{(n),s}_{\WA}(D)$, contributes equally to $D^+$ and $D^0$ decays.  In the flavor $SU(3)$ limit, the two contributions correspond to the sum of  isotriplet and isosinglet, and to the isosinglet contribution, respectively. 

\section{Results and Discussion}
Our first task will be to check whether the OPE at $O(1/m_c^3)$ describes the experimental data in a satisfactory way. In the pole mass scheme with $m_c=1.6$GeV, $r=0.005$, and $\mu_{\WA}=0.8$GeV we have 
\begin{align}
\Gamma &= \Gamma_0 \left[ 1- 0.72 \,\alpha_s -0.29\, \alpha_s^2 \beta_0 
-0.60 \,\mu^2_G - 0.20 \,\mu^2_\pi + 0.42 \,\rho^3_D + 0.38\, \rho^3_{LS} +
80 B_{\WA}^{(0)}\right]\,,\label{eq:Gammapole}\\
\braket{E} &= \braket{E}_0 \left[ 1- 0.03\,\alpha_s -0.03\, \alpha^2_s \beta_0 
-0.07 \,\mu^2_G + 0.20 \,\mu^2_\pi + 1.4 \,\rho^3_D + 0.29\, \rho^3_{LS} %-110 B_{\WA}^{(0)}
+135 \bar B_{\WA}^{(1)}\right]\,,\label{eq:E1pole}\\
\braket{E^2} &= \braket{E^2}_0 \left[ 1- 0.07 \,\alpha_s -0.05 \,\alpha^2_s \beta_0 
-0.14\,\mu^2_G +0.52 \,\mu^2_\pi + 3.5\,\rho^3_D + 0.66\, \rho^3_{LS} 
+204 \bar B_{\WA}^{(2)} \right]\,, \\
\sigma_E^2
 &= (\sigma_E^2)_0 \left[ 1- 0.09 \,\alpha_s -0.05 \,\alpha^2_s \beta_0 
-0.14 \,\mu^2_G +1.7 \,\mu^2_\pi + 9.4\,\rho^3_D + 1.4\, \rho^3_{LS} 
+ 641 \bar B_{\WA}^{(\sigma)}   \right]\,,
\label{eq:E2pole}
\end{align}
where $\beta_0=9$, $\Gamma_0$, $\braket{E}_0 $,  $\braket{E^2}_0 $, $(\sigma_E^2)_0$
are the tree-level results, and all coefficients are in GeV to the appropriate power. We immediately notice that the lepton moments receive smaller
perturbative corrections in the pole mass scheme than the total rate, while they are very sensitive to WA and $\rho_D^3$. Since the uncertainty on the value of $\rho_D^3$  to be employed in Eqs.~(\ref{eq:Gammapole}-\ref{eq:E2pole}) is 
sizable,  it follows from the previous section that the choice of $\mu_{\WA}$ has a strong impact on the final error on WA. Indeed, the simultaneous use of all three observables can reduce this ambiguity.
As for the $m_c$ dependence, 
since $\braket{E^n}$ scales  like $m_c^n$, the lowest moments are less sensitive to the value of the charm mass than the width,   which is proportional to $m_c^5$.
We have also introduced effective WA parameters $\bar B^{(1)}_{\WA}\simeq B^{(1)}_{\WA}-\frac35
B^{(0)}_{\WA}$, $\bar B^{(2)}_{\WA}\simeq B^{(2)}_{\WA}-\frac25
B^{(0)}_{\WA}$, $\bar B^{(\sigma)}_{\WA}\simeq B^{(0)}_{\WA}-3.8 B^{(1)}_{\WA}+3.2 B^{(2)}_{\WA} $.

In order  to avoid the well-known problems associated to the pole mass, semileptonic decays are usually described in terms of {\it threshold} heavy quark masses, like the kinetic mass, the PS, or the 1S mass, see \cite{ckmbook} for a review.  In the following we focus on the kinetic scheme \cite{kinetic}. For the charm mass and the non-perturbative OPE parameters
we use as initial inputs the results of a 
global fit to $B\to X_c \ell\nu$ and radiative moments \cite{Buchmuller:2005zv}, 
\begin{align}
m_c(1\rm GeV) &= 1.16(5) \mathrm{GeV}\,, & 
\mu_\pi^2 (1\rm GeV)&= 0.44(4) \mathrm{GeV}^2\,,\nonumber \\
\rho_{LS}^3 (1\rm GeV)&= -0.19(8) \mathrm{GeV}^3\,, & 
\rho_D^3(1\rm GeV) &= 0.19(2) \mathrm{GeV}^3\,.\nonumber
\end{align}
The above values refer to expectation values of local operators in the $B$ meson  in the kinetic scheme with the IR cutoff scale  $\mu_{kin} = 1$GeV. Up to power corrections, the $D$ meson expectation values can be identified with the $B$ meson ones, but $\mu_{kin}=1$GeV is  too high compared to the charm mass and we need to run the parameters to a lower IR scale 
above $\Lambda_{QCD} $.

We evolve the OPE parameters including the charm mass 
down to $\mu_{kin}=0.5$GeV using $O(\alpha_s^2)$ expressions. The expectation values   $\mu_G^2$ and $\rho^3_{LS}$ do not run with the kinetic scale. The running to such low scales induces significant uncertainties in the other parameters, of both perturbative and non-perturbative origin, which we estimate by varying the scale of $\alpha_s$. We finally adopt  $m_c(0.5{\rm GeV})= 1.40(7)$GeV, $\mu^2_\pi(0.5{\rm GeV})= 0.26(6)$GeV$^2$, $\rho_D^3(0.5{\rm GeV})=0.05(4)$GeV$^3$. Of course, the charm mass determination we use is not the most precise.  For instance, Ref.~\cite{Allison:2008xk} reports a very precise value in the $\overline{\rm MS}$ scheme which is consistent with \cite{Buchmuller:2005zv}. However, any scheme translation would increase the error significantly. In the end, we do not expect an improvement even in the case of the width. In addition, we assume that the large
uncertainty on $\rho^3_{LS}$ as extracted from the $B$ fit dominates over  the
perturbative and non-perturbative corrections to its value. In the case of $\mu_G^2$, on the other hand, we use a conservative approach with the central value of \cite{kolya} and a large error
$\mu_G^2 = 0.35(10)$GeV${}^2$.
For the strange quark we use its $\overline{\rm MS}$ definition,   $m_s(2\mathrm{GeV}) = 0.105(2)$GeV \cite{pdg}. We evaluate the corrections to the total rate and the first two leptonic moments at $\mu_{\WA} = 0.8 $GeV as
\begin{eqnarray}
 \Gamma_{kin} &=& 1.2(3) 10^{-13} \mathrm{GeV} \left\{1 
 + 0.23\,  \alpha_s  + 0.18\,  \alpha_s^2 \beta_0 - 0.79\,  \mu^2_G - 
 0.26 \mu^2_\pi + 1.45\,  \rho^3_D + 0.56 \rho^3_{LS}
  +120 B_{\WA}^{(0)} \right\},\\
\braket{E_\ell}_{kin} &=& 0.415(21) \mathrm{GeV} \left\{1 
+0.03\,  \alpha_s  + 0.02\,  \alpha_s^2 \beta_0 - 
 0.09\,  \mu^2_G + 0.26 \mu^2_\pi + 2.7 \rho^3_D + 0.44 \rho^3_{LS}
+203 \bar B_{\WA}^{(1)}\right\}\,,\\
\braket{E_\ell^2}_{kin} &=& 0.192(20) \mathrm{GeV}^2 \left\{ 1 
+0.001\,  \alpha_s  + 0.02 \, \alpha_s^2 \beta_0 - 
 0.18\,  \mu^2_G + 0.68 \mu^2_\pi + 6.6 \rho^3_D + 0.99 \rho^3_{LS}
+307\bar B_{\WA}^{(2)} \right\}\,,\\
\sigma^2_{E,kin} &=& 0.019(2) \mathrm{GeV}^2 \left\{ 1 
- 0.53\, \alpha_s - 
 0.17 \,\alpha_s^2 \beta_0 - 0.18 \mu^2_G + 2.2 \mu^2_\pi + 17 \rho^3_D + 
 2.1 \rho_{LS}^3
+961\bar B_{\WA}^{(\sigma)}\right\} \,,
\end{eqnarray}  
where all coefficients are in GeV to the appropriate power and the tree-level results include the  error due to their mass dependence.
In the first and the second leptonic energy moment, the dominant source of uncertainty is by far $\rho^3_D$, 
while in the total rate the charm quark mass uncertainty dominates.
In view of the accuracy relevant to our analysis, we can turn a blind eye to the bad convergence of the expansion for the width.
Varying all the input parameters as described above and  $\alpha_s$ by $\pm 20\%$ around  $\alpha_s(m_c)=0.36$ to account for higher order perturbative effects, 
we compare the theoretical expressions with eqs. (2,6) and
 we extract the following values of $B^{(0)}_{\WA}(D_q)$ and $\bar B^{(1,2,\sigma)}_{\WA}(D_q)$
\begin{align}\label{eq:B0}  
{ B}^{(0),s}_{\WA}(D^0) & = -0.001(3) \mathrm{GeV}^3 \,, & 
{ B}^{(0),s}_{\WA}{(D^+)} & = -0.001(3) \mathrm{GeV}^3 \,, & 
{ B^{(0),s}}_{\WA}{(D_s)} & = -0.002(3) \mathrm{GeV}^3 \,,    \\
{\bar B}^{(1),s}_{\WA}(D^0) & = -0.0001(6) \mathrm{GeV}^3 \,, & \label{eq:B1}
{\bar B}^{(1),s}_{\WA}{(D^+)} & = -0.0001(6) \mathrm{GeV}^3 \,, & 
{\bar B^{(1),s}}_{\WA}{(D_s)} & = -0.0001(6) \mathrm{GeV}^3 \,,    \\
  {\bar B^{(2),s}}_{\WA}{(D^0)} & = -0.0001(10) \mathrm{GeV}^3 \,, &
  {\bar B^{(2),s}}_{\WA}{(D^+)} & = -0.0002(10) \mathrm{GeV}^3 \,,  &  
{\bar B^{(2),s}}_{\WA}{(D_s)} & = -0.0001(10) \mathrm{GeV}^3 \,,    \\
    {\bar B^{(\sigma),s}}_{\WA}{(D^0)} & = -0.0000(7) \mathrm{GeV}^3\,,& 
  {\bar B^{(\sigma),s}}_{\WA}{(D^+)} & = -0.0000(7) \mathrm{GeV}^3\,, &
{\bar B^{(\sigma),s}}_{\WA}{(D^s)} & = 0.0001(10) \mathrm{GeV}^3
\,. \label{eq:Bsigma}
\end{align}
At the estimated theoretical precision, all the extracted WA contributions are consistent with zero. The obvious implication is  that the OPE describes all the data reasonably well.
  We also repeated the exercise at different kinetic scales, finding in general similar results, despite the high sensitivity to accidental cancellations.   For larger $\mu_{kin}$
the size of perturbative corrections to the rate and their associated error  increase rapidly as noticed in \cite{Kamenik:2009ze}. However, the moments are less affected and one obtains consistent WA estimates for kinetic scales as large as $\mu_{kin}=0.8$ GeV with similar errors.
We have also repeated our analysis using the 1S scheme for the charm mass \cite{1S} and $\mu_{kin}=0$ for the other OPE parameters. The apparent convergence of the perturbative expansion improves for the rate and variance, but the estimates and errors of the WA contributions are very similar.

Eqs.(\ref{eq:B0}-\ref{eq:Bsigma}) show no departure  from the $SU(3)$ flavor symmetry, but
the errors are strongly correlated among different mesons. On the other hand,
in the ratios of $D_s$ and $D^0$ semileptonic widths and leptonic
moments, the dominant $m_c$ and all the perturbative  corrections cancel, while only isotriplet WA and $SU(3)$ breaking contributions to the OPE parameters  remain. We estimate
the latter conservatively as 20\% of the central values and add the
resulting errors linearly to account for possible correlations. Comparing
these expressions to the experimental values we obtain $\Delta B_{\WA}^{(0),s} \equiv
B_{\WA}^{(0),s}(D_s)-B_{\WA}^{(0),s}(D^0)=
-0.0014(12)(5)\mathrm{ GeV}^3$ from the ratio of rates, $\Delta \bar B_{\WA}^{(1),s} =
0.0000(3)(1) \mathrm{GeV}^3$ from the ratio of the first moments, and $\Delta \bar B_{\WA}^{(2),s} = 0.0000(4)(2)
\mathrm{GeV}^3$ from the ratio of the second leptonic energy moments, where we show the theoretical and experimental uncertainty contributions in the first and the second brackets respectively. We observe a mild indication
of a nonzero negative isotriplet $B_{\WA}^{(0),s}$ contribution.
This indication is however not really significant at  present. 

So far we have parameterized the WA contributions to the individual moments, and we have learned that the OPE gives a satisfactory description of the data even in the absence of WA, with the possible exception of $\Delta B_{\WA}^{(0)s}$.
As already mentioned in the previous Section, we expect WA to be concentrated at the end-point, in which case we can combine the $D^{0,+}$ results
 from eqs.~(\ref{eq:B1},\ref{eq:Bsigma}) to
obtain the isosinglet  $B_{\WA}^{(0)s} = -0.0003(15)$ GeV${}^3$.
In the worst case, the reduction in sensitivity due to the WA smearing functions shown in Fig.~2  
is about  60\%, which leads to  
\begin{equation}
B_{\WA}^{s} = -0.0003(25)\mathrm{GeV}^3\,, \label{eq:charm}
\end{equation}
where we have combined the dilution error linearly, without changing the central value. We regard this as sufficiently conservative. We can also extract $\Delta B_{\WA}^{(0),s}$ from 
$\Delta \bar B_{\WA}^{(1,2),s}$: in the absence of WA smearing
they both give the same result, $\Delta B_{\WA}^{(0),s}=0.0000(7)(3)$, while the worst dilution scenario gives $\Delta B_{\WA}^{(0),s}=0.0000(12)(3)$, which is  compatible 
with the value extracted from the ratio of widths.

\begin{figure}
  \includegraphics[height=.3\textheight]{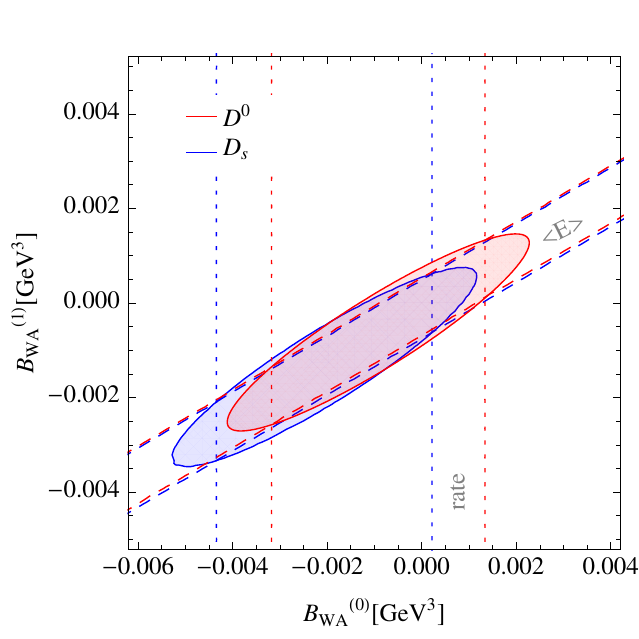}
  \includegraphics[height=.3\textheight]{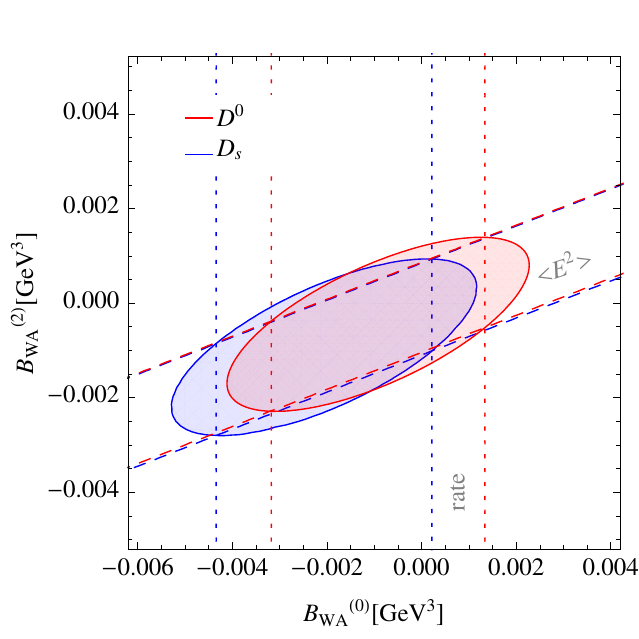}
  \caption{\label{fig:combined} Combined constraints on $B^{(i),s}_{WA}$ from the total semileptonic rates of $D^0$ and $D_s$ and the first two leptonic energy moments.}
\end{figure}

We can also look for indications of WA dilution by extracting $B^{(0,1,2)}_{\WA}$ from a combined fit to the moments and the total rates (recall that $\bar B_{\WA}^{(1,2)}$ are linear combinations of $B^{(0,1,2)}_{\WA}$). The results  for $D^0$ and $D_s$ decays are shown in figure \ref{fig:combined}
and are consistent with $B^{(0)}_{\WA} \approx B^{(1)}_{\WA} \approx B^{(2)}_{\WA}$, but the errors are too large to draw a conclusion concerning WA dilution.

Finally we study the correlation between $\rho^3_D$ and WA  by considering $\rho^3_D$
a free parameter in the fit. In this way, the dominant source of uncertainty in the moments is removed. From the variance, first and second moments of $D^0$ we find the following  constraints (the results from $D^+$ and $D_s$ data are consistent)
\begin{align}
\bar B^{(\sigma),s}_{\WA} + 0.017  \rho^3_D & =  0.0009(2) \mathrm{GeV^3}\,, & \bar B^{(2),s}_{\WA} + 0.021 \rho^3_D & = 0.0010(5) \mathrm{GeV^3}\,, & \bar B^{(1),s}_{\WA} + 0.014 \rho^3_D & = 0.0006(4) \mathrm{GeV^3}\,.  \nonumber
\end{align}
Assuming no WA dilution one then obtains in each of these cases
\begin{align}
B^{s}_{\WA} + 0.043  \rho^3_D & =  0.0021(5) \mathrm{GeV^3}\,, & B^{s}_{\WA} + 0.035 \rho^3_D & = 0.0016(9) \mathrm{GeV^3}\,, & B^{s}_{\WA} + 0.033 \rho^3_D & = 0.0014(10) \mathrm{GeV^3}\,, \nonumber
\end{align}
and it is remarkable that they all agree quite well.  This may be viewed as a mild indication that there is no significant WA dilution, or possibly that the dilution is  similar in the three cases. Indeed, assuming vanishing WA at $\mu_{\WA}=0.8$GeV all the moments  can be reproduced by  $\rho^3_D(0.5\mathrm{GeV})=0.05(1)$GeV${}^3$.

In order to connect to $B\to X_u \ell\nu$ we refer to the WA matrix elements determined in charm decays as $B^{cq}_{\WA}$ and consider their relation to those relevant in $B$ semileptonic decays, $B^{bq}_{\WA}$, taking into account eq.~(\ref{eq:me}).
In the heavy quark limit, $f_P \sim m_P^{-1/2}$ \cite{Grinstein:1992qt} so that $B_{P,i}$ scale as constants with heavy quark mass, but
recent lattice results give $f_D \approx 0.21$~GeV, $f_B\approx 0.20$~GeV \cite{Lubicz:2008am}. We also neglect any evolution of the WA operators and 
write
\begin{equation}\label{eq:25}
B^{bq}_{\WA}(\mu_{\WA}) =  \frac{m_B f_B^2}{m_D f_D^2} B^{cq}_{\WA}  (\mu_{\WA})\,.
\end{equation}
The parametric enhancement due to meson masses and decay constants is a significant factor of $2.5$.
Due to finite heavy quark masses, one might also expect additional power corrections, which spoil the exact scaling of $B_{\WA}$ between the $D$ and $B$ sectors. 
From  eqs.~(\ref{eq:charm},\ref{eq:25}) we derive  a bound
\begin{equation}
|B^{b}_{\WA}(\mu_{\WA}=0.8\mathrm{GeV})| \lesssim 0.006\mathrm{GeV}^{3}\,,
\label{eq:bound}
\end{equation}
which holds for the non-valence contributions, although we stress that 
the data seem to prefer even smaller values.  The valence contribution is
 more constrained from the ratio of the $D_s$ and $D^0$ rates and leptonic moments, from which we 
expect
 \begin{equation}
-0.004\,\mathrm{GeV}^{3}\lesssim \Delta B^{b}_{\WA}(\mu_{\WA}=0.8\mathrm{GeV}) \lesssim 0.002\,\mathrm{GeV}^{3}.
\end{equation}
The  bounds lead to a maximum  2\%  WA correction  to the total rate of $B\to X_u \ell \nu$. In turn, this translates into an uncertainty of 1\%  on $|V_{ub}|$ extracted from the total rate and from the most inclusive experimental analyses, like those that  involve  a lower cut on the invariant hadronic mass. 
Our bound on the WA expectation value improves on previous estimates.
In \cite{Vub}, for instance, the maximum value allowed for $B^{b}_{\WA}(\mu_{\WA}=1\mathrm{GeV})$ was as high as $0.020\mathrm{GeV}^{3}$,
or $B^{b}_{\WA}(\mu_{\WA}=0.8\mathrm{GeV})=0.018\mathrm{GeV}^{3}$.

Our analysis is compatible with the one of ref.~\cite{Bigi:2009ym} for the valence contribution, 
although we allow for a larger SU(3) violation. The analysis of ref.~\cite{Ligeti:2010vd} gives a determination of both valence and non-valence WA  from the $D$ semileptonic widths only. In our notation the results of ref.~\cite{Ligeti:2010vd} correspond
to a valence contribution $\Delta B^{cs}_{\WA}=-0.0015(9){\rm GeV}^3$ and to an isosinglet contribution $B^{cs}_{\WA}=0.0036(5){\rm GeV}^3$.
The valence contribution is perfectly compatible with our estimate from the widths only. 
The non-valence contribution, on the other hand, has a tiny error and should be compared with $B^{cs}_{\WA}=-0.001(3){\rm GeV}^3$ that we extract from the widths only, see eq.~(19).
The difference is presumably due to various sources: $i)$ the charm mass that we employ is 
higher than that employed in \cite{Ligeti:2010vd}: we start with the result of the global fit 
\cite{Buchmuller:2005zv}, which is in excellent agreement with other $m_c$ determinations like charmonium sum rules, and evolve it to lower  $\mu_{kin}$; 
$ii)$ our OPE parameters (taken from  \cite{Buchmuller:2005zv} and then evolved to lower  $\mu_{kin}$)   are  different from those 
of ref.~\cite{Ligeti:2010vd}, which  may lead to sizable differences in charm but not in bottom decays; $iii)$ the method of  \cite{Ligeti:2010vd} implies the use of $\alpha_s(m_b)$,  and might underestimate the perturbative corrections. The discrepancy between the two determinations provides indeed additional motivation for using the moments to constrain WA, as they 
are less sensitive to power and perturbative corrections.
In any case, we find  remarkable that our determination from the moments is compatible with that from the widths and relatively stable wrt changes in the definition of the charm mass or in $\mu_{kin}$. 
Notice also that despite the larger value of  $B^{cs}_{\WA}$ they find, the authors of \cite{Ligeti:2010vd} end up with an estimate for the WA effects in  the $B\to X_u \ell \nu$ rate which is very similar to ours, once the enhancement related to eq.~(\ref{eq:25}) is taken into account.
 This is because our large uncertainty  almost covers the discrepancy in the central values.

\section{Summary}
We have performed an analysis of inclusive  semileptonic $D$ decays using the Heavy Quark Expansion. In addition to the total widths, we have used the Cleo data on the lepton energy spectra  to compute the first few moments. The latter are 
quite sensitive probes of possible  Weak Annihilation contributions, both in  its isosinglet and isotriplet  
components, and determine very precisely a linear combination of the expectation values of the Darwin and WA operators.
We have shown that the extraction of WA from the moments depends to the way WA is distributed in the lepton energy spectrum and we have taken this effect into account in our error estimates. 

 Our analysis of Cleo data shows no evidence for Weak Annihilation, i.e.\ the OPE describes
 well the experimental results even in the absence of WA.
  There is a mild indication 
of non-zero valence WA in the ratio of the $D_s$ and $D^0$ widths, but it does not seem to be supported by the ratio of the lepton moments  in $D_s$ and $D^0$ decays. A possible explanation involves  sizable $SU(3)$ breaking in the matrix elements of the power suppressed operators and/or a WA contribution which is broadly distributed over the leptonic spectrum.

We derive an upper limit on both valence and non-valence WA components, 
which allows us to put a  bound of 2\% on their relevance in the $B\to X_u \ell\nu$ decay rate and even less for the isotriplet component. 
We look forward to the individual measurements of the $B^+$ and $B^0$ charmless inclusive semileptonic decays, which could test our conclusions regarding the suppression of valence WA in these decays.

Finally we note that the hadronic mass and $q^2$ moments in inclusive semileptonic decays of heavy quarks are quite sensitive to non-perturbative contributions, see  e.g.\  \cite{bumom}. 
Their measurement at Cleo-c  or at the recently started BESIII experiment~\cite{BESIII} 
might help us disentangle the various non-perturbative effects.

\begin{acknowledgments}

  We thank Ikaros Bigi and Nikolai Uraltsev for enlightening discussions and various comments on the manuscript, the Galileo Galilei Institute for Theoretical Physics where this work has been finalized during the workshop {\it Indirect searches for new physics at the time of LHC}, and  INFN for partial support during the completion of this
work.  JFK acknowledges useful discussions with Thomas Mannel and is supported in part by the European Commission RTN  network, Contract No. MRTN-CT-2006-035482 (FLAVIAnet) and by the Slovenian Research Agency. PG is supported in part by a EU's Marie-Curie Research Training Network under contract
MRTN-CT-2006-035505 (HEPTOOLS).

\end{acknowledgments}


\begin{thebibliography}{99}                                                                                                

\bibitem{UTfit}M.~Bona {\it et al.}  [UTfit Collaboration],
  %``An Improved Standard Model Prediction Of BR(B -> tau nu) And Its
  %Implications For New Physics,''
  arXiv:0908.3470 [hep-ph], see also {\tt http://www.utfit.org/};
  %%CITATION = ARXIV:0908.3470;%%
for recent $|V_{ub}|$ averages see {\tt http://www.slac.stanford.edu/xorg/hfag/}.

\bibitem{ckmbook} M.~Antonelli {\it et al.},
  %``Flavor Physics in the Quark Sector,''
  arXiv:0907.5386 [hep-ph].
  %%CITATION = ARXIV:0907.5386;%%

\bibitem{OPE}
J. ~Chay, H.~Georgi, and B.~Grinstein, 
Phys.\ Lett.\ B{\bf 247} (1990) 399;
I.~I.~Y.~Bigi, N.~G.~Uraltsev, and A.~I.~Vainshtein, 
Phys.\ Lett.\ B{\bf 293} (1992) 430 [arXiv:hep-ph/9207214] [Erratum, ibid. B {\bf 297} (1993) 477];
I.~I.~Y.~Bigi, M.~A.~Shifman, N.~G.~Uraltsev, and A.~I.~Vainshtein, 
Phys.\ Rev.\ Lett.\ {\bf 71} (1993) 496 [arXiv:hep- ph/9304225];
A.~V.~Manohar and M.~B.~Wise, 
Phys.\ Rev.\ D{\bf 49} (1994) 1310  [arXiv:hep-ph/9308246].

  \bibitem{Bigi:1993bh}
  I.~I.~Y.~Bigi and N.~G.~Uraltsev,
  %``Weak annihilation and the endpoint spectrum in semileptonic B decays,''
  Nucl.\ Phys.\  B {\bf 423}, 33 (1994)
  [arXiv:hep-ph/9310285].
  %%CITATION = NUPHA,B423,33;%%

\bibitem{Bigi:1997dn}
 R.~D.~Dikeman and N.~G.~Uraltsev,
  %``Key distributions for charmless semileptonic B decay,''
  Nucl.\ Phys.\  B {\bf 509}, 378 (1998)
  [arXiv:hep-ph/9703437];
  %%CITATION = NUPHA,B509,378;%%
  I.~I.~Y.~Bigi, R.~D.~Dikeman and N.~Uraltsev,
  %``The hadronic recoil mass spectrum in semileptonic B decays and  extracting
  %|V(ub)| in a model-insensitive way,''
  Eur.\ Phys.\ J.\  C {\bf 4}, 453 (1998)
  [arXiv:hep-ph/9706520].
  %%CITATION = EPHJA,C4,453;%%
  
\bibitem{QCDSRHQ}
  M.~S.~Baek, J.~Lee, C.~Liu and H.~S.~Song, 
  Phys.\ Rev.\ D {\bf 57} (1998) 4091 [arXiv:hep-ph/9709386]; 
  H.~Y.~Cheng and K.~C.~Yang, 
  Phys. Rev. D {\bf 59} (1999) 014011 [arXiv:hep-ph/9805222].


  %\cite{DiPierro:1998ty}
\bibitem{DiPierro:1998ty}
  M.~Di Pierro and C.~T.~Sachrajda  [UKQCD Collaboration],
  %``A lattice study of spectator effects in inclusive decays of B mesons,''
  Nucl.\ Phys.\  B {\bf 534} (1998) 373
  [arXiv:hep-lat/9805028].
  %%CITATION = NUPHA,B534,373;%%

  %\cite{Becirevic:2001fy}
\bibitem{Becirevic:2001fy}
  D.~Becirevic,
  %``Theoretical progress in describing the B meson lifetimes,''
  arXiv:hep-ph/0110124.
  %%CITATION = HEP-PH/0110124;%%

\bibitem{Pirjol:1998ur}
  D.~Pirjol and N.~Uraltsev,
  %``Four-fermion heavy quark operators and light current amplitudes in  heavy
  %flavor hadrons,''
  Phys.\ Rev.\  D {\bf 59} (1999) 034012
  [arXiv:hep-ph/9805488].
  %%CITATION = PHRVA,D59,034012;%%
\bibitem{Uraltsev:1999rr}
  N.~Uraltsev,
  %``Theoretical uncertainties in Gamma(sl)(b --> u),''
  Int.\ J.\ Mod.\ Phys.\  A {\bf 14} (1999) 4641
  [arXiv:hep-ph/9905520].
  %%CITATION = IMPAE,A14,4641;%%

\bibitem{Bigi:1993ey}
  I.~I.~Y.~Bigi and N.~G.~Uraltsev,
  %``D(s) Lifetime, m(b), m(c) and $|$V(cb)$|$ in the heavy quark expansion,''
  Z.\ Phys.\  C {\bf 62}, 623 (1994)
  [arXiv:hep-ph/9311243].
  %%CITATION = ZEPYA,C62,623;%%

%\cite{Voloshin:2001xi}
\bibitem{Voloshin:2001xi}
  M.~B.~Voloshin,
  %``Nonfactorization effects in heavy mesons and determination of |V(ub)|  from
  %inclusive semileptonic B decays,''
  Phys.\ Lett.\  B {\bf 515}, 74 (2001)
  [arXiv:hep-ph/0106040].
  %%CITATION = PHLTA,B515,74;%%
  
  %\cite{Becirevic:2008us}
\bibitem{Becirevic:2008us}
  D.~Becirevic, S.~Fajfer and J.~F.~Kamenik,
  %``On the matrix elements of dB=0 operators in the heavy meson decay widths,''
  Phys.\ Lett.\  B {\bf 671}, 66 (2009)
  [arXiv:0804.1750 [hep-ph]].
  %%CITATION = PHLTA,B671,66;%%
    
  %\cite{Bigi:2009ym}
\bibitem{Bigi:2009ym}
  I.~Bigi, T.~Mannel, S.~Turczyk and N.~Uraltsev,
  %``The Two Roads to 'Intrinsic Charm' in B Decays,''
  arXiv:0911.3322 [hep-ph].
  %%CITATION = ARXIV:0911.3322;%%

%\cite{Ligeti:2010vd}
\bibitem{Ligeti:2010vd}
  Z.~Ligeti, M.~Luke and A.~V.~Manohar,
  %``Constraining weak annihilation using semileptonic D decays,''
  arXiv:1003.1351 [hep-ph].
  %%CITATION = ARXIV:1003.1351;%%
    %\cite{:2009pu}

\bibitem{:2009pu}
  D.~M.~Asner et al.  [The CLEO Collaboration],
  %``Inclusive semileptonic decays of charm and charmed-strange mesons,''
  arXiv:0912.4232 [hep-ex].
  %%CITATION = ARXIV:0912.4232;%%


  \bibitem{Blok:1994cd}
  B.~Blok, R.~D.~Dikeman and M.~A.~Shifman,
  %``Calculation of 1/m(c)**3 terms in the total semileptonic width of D
  %mesons,''
  Phys.\ Rev.\  D {\bf 51} (1995) 6167
  [arXiv:hep-ph/9410293].
  %%CITATION = PHRVA,D51,6167;%%
  
  \bibitem{fD}
  A.~Bazavov {\it et al.}  [Fermilab Lattice and MILC Collaborations],
  %``The Ds and D+ Leptonic Decay Constants from Lattice QCD,''
  PoS {\bf LAT2009} (2009) 249
  [arXiv:0912.5221 [hep-lat]].
  %%CITATION = POSCI,LAT2009,249;%%
  
\bibitem{Leibovich:2002ys}
  A.~K.~Leibovich, Z.~Ligeti and M.~B.~Wise,
  %``Enhanced subleading structure functions in semileptonic B decay,''
  Phys.\ Lett.\  B {\bf 539} (2002) 242
  [arXiv:hep-ph/0205148].
  %%CITATION = PHLTA,B539,242;%%

%\cite{Pak:2008qt}
\bibitem{Pak:2008qt}
  A.~Pak and A.~Czarnecki,
  %``Mass effects in muon and semileptonic b -> c decays,''
  Phys.\ Rev.\ Lett.\  {\bf 100}, 241807 (2008)
  [arXiv:0803.0960 [hep-ph]] 
  %%CITATION = PRLTA,100,241807;%%
and K. Melnikov,
 %``O(\alpha_s^2) corrections to semileptonic decay b \to c l \bar \nu_l,''
  Phys.\ Lett.\  B {\bf 666}, 336 (2008)
  [arXiv:0803.0951 [hep-ph]].
  %%CITATION = PHLTA,B666,336;%%


\bibitem{Aquila:2005hq}
  V.~Aquila, P.~Gambino, G.~Ridolfi and N.~Uraltsev,
  %``Perturbative corrections to semileptonic b decay distributions,''
  Nucl.\ Phys.\  B {\bf 719} (2005) 77
  [arXiv:hep-ph/0503083].
  %%CITATION = NUPHA,B719,77;%%


%\cite{Gremm:1996df}
\bibitem{Gremm:1996df}
  M.~Gremm and A.~Kapustin,
  %``Order $1/m_b~3$ corrections to inclusive semileptonic $B$ decay,''
  Phys.\ Rev.\  D {\bf 55} (1997) 6924
  [arXiv:hep-ph/9603448].
  %%CITATION = PHRVA,D55,6924;%%
  
  %\cite{Dassinger:2006md}
\bibitem{Dassinger:2006md}
  B.~M.~Dassinger, T.~Mannel and S.~Turczyk,
  %``Inclusive semi-leptonic B decays to order 1/m(b)**4,''
  JHEP {\bf 0703}, 087 (2007)
  [arXiv:hep-ph/0611168].
  %%CITATION = JHEPA,0703,087;%%


%\cite{Gambino:2008fj}
\bibitem{Gambino:2008fj}
  P.~Gambino and P.~Giordano,
  %``Normalizing inclusive rare B decays,''
  Phys.\ Lett.\  B {\bf 669}, 69 (2008)
  [arXiv:0805.0271 [hep-ph]].
  %%CITATION = PHLTA,B669,69;%%
  
  %\cite{Czarnecki:1994pu}
\bibitem{Czarnecki:1994pu}
  A.~Czarnecki and M.~Jezabek,
  %``Distributions of leptons in decays of polarized heavy quarks,''
  Nucl.\ Phys.\  B {\bf 427}, 3 (1994)
  [arXiv:hep-ph/9402326].
  %%CITATION = NUPHA,B427,3;%%
   
\bibitem{bumom}
  P.~Gambino, G.~Ossola and N.~Uraltsev,
  %``Hadronic mass and q**2 moments of charmless semileptonic B decay
  %distributions,''
  JHEP {\bf 0509} (2005) 010
  [arXiv:hep-ph/0505091].
  %%CITATION = JHEPA,0509,010;%%

\bibitem{Vub}   
    P.~Gambino, P.~Giordano, G.~Ossola and N.~Uraltsev,
  %``Inclusive semileptonic B decays and the determination of |V_ub|,''
  JHEP {\bf 0710} (2007) 058
  [arXiv:0707.2493 [hep-ph]].
  %%CITATION = JHEPA,0710,058;%%
  
  \bibitem{kinetic}
I.~I.~Y.~Bigi, M.~A.~Shifman, N.~Uraltsev and A.~I.~Vainshtein,
%``High power n of m(b) in beauty widths and n = 5 $\to$ infinity limit,''
Phys.\ Rev.\ D {\bf 56} (1997) 4017
[arXiv:hep-ph/9704245]
%%CITATION = HEP-PH 9704245;%%
and 
%``Sum rules for heavy flavor transitions in the SV limit,''
Phys.\ Rev.\ D {\bf 52} (1995) 196
[arXiv:hep-ph/9405410].
%%CITATION = HEP-PH 9405410;%%

  
%\cite{Buchmuller:2005zv}
\bibitem{Buchmuller:2005zv}
  O.~Buchmuller and H.~Flacher,
  %``Fits to moment measurements from B --> X/c l nu and B --> X/s gamma  decays
  %using heavy quark expansions in the kinetic scheme,''
  Phys.\ Rev.\  D {\bf 73} (2006) 073008
  [arXiv:hep-ph/0507253], updated for the PDG 09 web update, {\tt{http://www.slac.stanford.edu/xorg/hfag2/semi/winter09}};
  %%CITATION = PHRVA,D73,073008;%%
  P.~Gambino and N.~Uraltsev,
  %``Moments of semileptonic B decay distributions in the 1/m(b) expansion,''
  Eur.\ Phys.\ J.\  C {\bf 34} (2004) 181
  [arXiv:hep-ph/0401063].
  %%CITATION = EPHJA,C34,181;%%
  

\bibitem{Allison:2008xk}
  I.~Allison {\it et al.}  [HPQCD Collaboration],
  %``High-Precision Charm-Quark Mass from Current-Current Correlators in Lattice
  %and Continuum QCD,''
  Phys.\ Rev.\  D {\bf 78} (2008) 054513
  [arXiv:0805.2999 [hep-lat]].
  %%CITATION = PHRVA,D78,054513;%%

\bibitem{kolya}
  N.~Uraltsev,
  %``On the chromomagnetic expectation value mu(G)**2 and higher power
  %corrections in heavy flavor mesons,''
  Phys.\ Lett.\  B {\bf 545} (2002) 337
  [arXiv:hep-ph/0111166].
  %%CITATION = PHLTA,B545,337;%%
 
  \bibitem{pdg}
   C. ~Amsler et al. [Particle Data Group], 
   Phys.\ Lett.\ B{\bf 667}, 1 (2008) and 2009 partial update for the 2010 edition.

\bibitem{1S}  A.~H.~Hoang, Z.~Ligeti and A.~V.~Manohar,
  %``B decays in the Upsilon expansion,''
  Phys.\ Rev.\  D {\bf 59}, 074017 (1999)
  [arXiv:hep-ph/9811239].

\bibitem{Kamenik:2009ze}
  J.~F.~Kamenik,
  %``Theory of Semileptonic Charm Decays,''
  arXiv:0909.2755 [hep-ph].
  %%CITATION = ARXIV:0909.2755;%%
  
%  \cite{Grinstein:1992qt}
\bibitem{Grinstein:1992qt}
  B.~Grinstein, E.~E.~Jenkins, A.~V.~Manohar, M.~J.~Savage and M.~B.~Wise,
  %``Chiral Perturbation Theory for $f_{D_S}/f_D$ and $B_{B_S}/B_{B}$,''
  Nucl.\ Phys.\  B {\bf 380}, 369 (1992)
  [arXiv:hep-ph/9204207].
  

\bibitem{Lubicz:2008am}
  V.~Lubicz and C.~Tarantino,
  %``Flavour physics and Lattice QCD: averages of lattice inputs for the
  %Unitarity Triangle Analysis,''
  Nuovo Cim.\  {\bf 123B}, 674 (2008)
  [arXiv:0807.4605 [hep-lat]].
  %%CITATION = NUCIA,123B,674;%%
      
  \bibitem{BESIII}
D.~M.~Asner {\it et al.},
  %``Physics at BES-III,''
  arXiv:0809.1869 [hep-ex].
  %%CITATION = ARXIV:0809.1869;%%
  
        
\end{thebibliography}
\end{document}